\documentclass[runningheads]{llncs}
\usepackage[T1]{fontenc}

\usepackage{graphicx}

\usepackage{amsmath,amssymb,amsfonts}
\usepackage{algorithmic}
\usepackage{subfigure}
\usepackage{float}
\usepackage{textcomp}
\usepackage{xcolor}
\usepackage{cite}

\begin{document}
\title{Towards Sustainable Precision: Machine Learning for Laser Micromachining Optimization}

\titlerunning{Machine Learning for Laser Micromachining Optimization}

\author{Luis Correas-Naranjo\inst{1}\and
Miguel Camacho-Sánchez\inst{1} \and Laëtitia Launet\inst{1} \and Milena Zuric\inst{2} \and
Valery Naranjo\inst{1}}
\authorrunning{Correas-Naranjo et al.}

\institute{
Instituto Universitario de Investigación en Tecnología Centrada en el Ser Humano,\\
HUMAN-tech, Universitat Politècnica de València, Valencia, Spain
\\ \email{vnaranjo@upv.es}\and
Fraunhofer Institute for Laser Technology ILT\\
Steinbachstraße 15, 52074 Aachen, Germany
}
\maketitle
\begin{abstract}
In the pursuit of sustainable manufacturing, ultra-short pulse laser micromachining stands out as a promising solution while also offering high-precision and qualitative laser processing. However, unlocking the full potential of ultra-short pulse lasers requires an optimized monitoring system capable of early detection of defective workpieces, regardless of the preprocessing technique employed. While advances in machine learning can help predict process quality features, the complexity of monitoring data necessitates reducing both model size and data dimensionality to enable real-time analysis. To address these challenges, this paper introduces a machine learning framework designed to enhance surface quality assessment across diverse preprocessing techniques. To facilitate real-time laser processing monitoring, our solution aims to optimize the computational requirements of the machine learning model. Experimental results show that the proposed model not only outperforms the generalizability achieved by previous works across diverse preprocessing techniques but also significantly reduces the computational requirements for training. Through these advancements, we aim to establish the baseline for a more sustainable manufacturing process.
\end{abstract}
\vspace{-1em}
\section{Introduction}

\vspace{-0.6em}

In recent years, the evolution of laser machining techniques has transformed the manufacturing field.
Among these innovative methods, ultra-short pulsed (USP) laser structuring has positioned itself as a competitive approach for achieving precise and high-quality laser micromachining across a wide range of materials.

As sustainability is becoming a priority in modern manufacturing practices, USP laser processing offers a more sustainable alternative to traditional manufacturing processes, with the absence of process or waste chemicals. Incorporating online monitoring into the laser processing can also help optimize the entire process, to drastically decrease wasted workpieces while increasing machine productivity.

Achieving an effective process monitoring of laser processing that minimizes workpiece waste requires a real-time system to enhance the quality assurance of micromachining. In recent years, the advent of machine learning (ML) methods has facilitated the application of innovative automatic algorithms to a wide variety of fields, including manufacturing \cite{knaak2018machine, amini2018process, mcdonnell2021machine}. Knaak et al. \cite{knaak2018machine} introduced a random forest algorithm for assessing laser welding quality with data from two cameras, while Amini et. al \cite{amini2018process} monitored selective laser melting on previously selected K-means clusters, and showed that more than 50 parameters impacted the printing quality. In a different approach, McDonnell et al. \cite{mcdonnell2021machine} used neural networks to optimize laser parameters before laser processing.
Despite these advancements, monitoring USP laser processing remains a significant challenge due to the demanding requirements of high spatial resolutions and processing speeds \cite{zuric2019multi}, hence the need to optimize ML models to enable real-time implementation.

Simultaneously, current algorithms for USP laser process monitoring struggle to achieve promising results across inherently different preprocessing techniques such as milling, grinding, or polishing. In \cite{zuric2023}, Leyendecker et al. developed an ML solution to predict surface roughness, but results showed to be strongly dependent on workpiece properties and preprocessing techniques. This sensitivity to preprocessing techniques underscores the need for a versatile and effective system capable of meeting the diverse demands of micromachining processes.

In light of these challenges, this paper proposes an ML-based approach to address the challenges in USP laser micromachining, aiming to enhance results across multiple preprocessing techniques. Considering roughness as a key feature that reflects the overall process quality, the proposed approach focuses on predicting the final surface roughness after laser structuring, utilizing initial laser parameters and photodiode sensor data as inputs. Recognizing the high dimensionality of sensor data, this research also aims to strategically reduce input features to optimize monitoring ML models. This not only tackles the computational cost associated with training but also facilitates the application of lighter ML models, aiming for efficient real-time monitoring of the USP laser process.

\vspace{-1em}
\section{Material}
\label{sec:material}

\begin{figure}[!htb]
    \centering
    \includegraphics[width=0.8\textwidth]{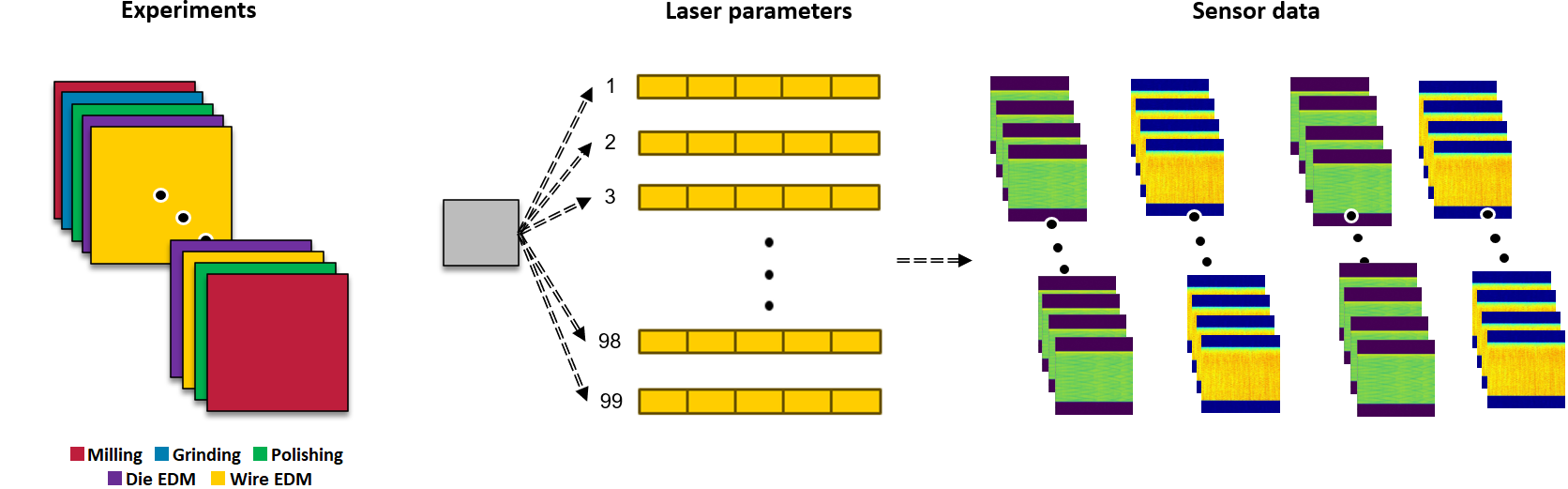} 
    \vspace{-0.8em}
    \caption{Data structure for the different experiments leveraged in this project. (Left) Set of workpiece experiments with diverse preprocessing techniques; (Center) For each workpiece, 99 sub-experiments are carried out with varying laser parameters; (Right) Resulting sensor data for each available sub-experiment, consisting of 5684 layers in total per experiment.}
    \label{fig:data}
    \vspace{-1.2em}
\end{figure}

\vspace{-0.4em}
In this work, we leveraged a multi-sensor dataset for process monitoring provided by the Fraunhofer Institute for Laser Technology (ILT). As depicted in Fig.\ref{fig:data}, the dataset comprises 60 experiments conducted on one of the lasers, each associated with a different workpiece and diverse preprocessing techniques, including milling, grinding, polishing, wire-EDM, and die-EDM. Within each experiment, 99 distinct sub-experiment samples were generated, each gathering information about laser parameters, initial and final roughness, along with sensor emission data for every layer. This sensor data was obtained from four photodiodes strategically mounted around the focusing lens, capturing infrared (IR), acoustic, laser reflection, and visible light (VIS) emissions. To generate the dataset, laser parameters were defined prior to processing for the samples of each experiment, thus resulting in a set of 99 parameters that were consistently used across all experiments. Specifically, these varying parameters correspond to the following: pulses per burst, pulse fluence($J/cm^2$), laser power ($W$), and number of layers (laser scans over a particular area).

Building on this dataset, our approach aims to predict the final roughness of a sample regardless of the preprocessing applied (e.g., milling, grinding, etc.) before surface processing with the laser, while reducing input features for optimized monitoring.

\vspace{-1.6em}
\section{Methods}
\label{sec:methods}
\vspace{-1.5em}

\begin{figure*}[h!btp]
    \centering
    \includegraphics[width=\textwidth]{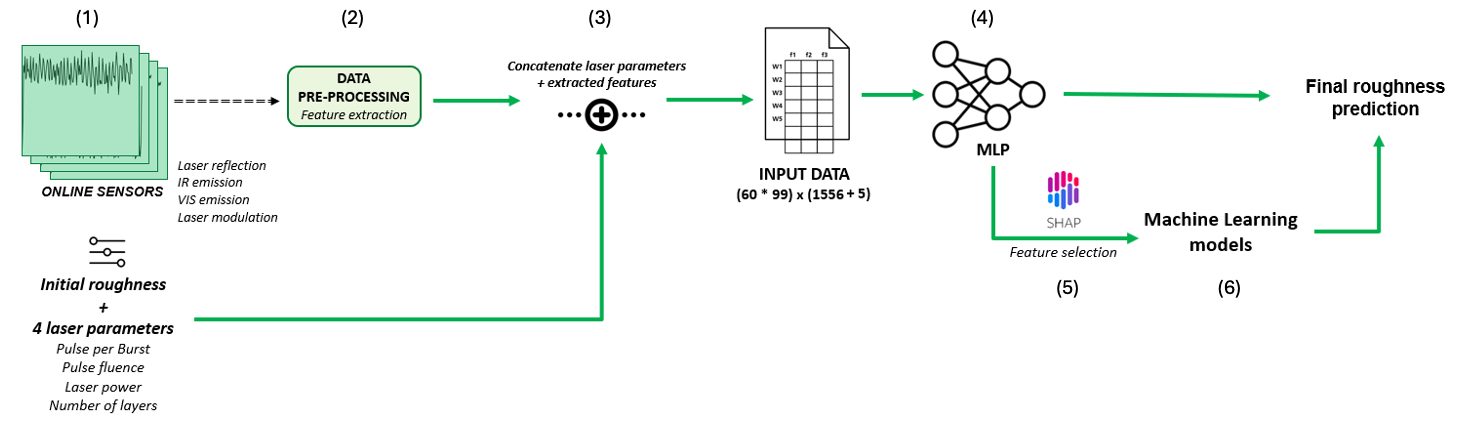}
    \vspace{-0.7em}
    \caption{Overall ML-based pipeline for optimized surface roughness prediction, (1) leverage data from the laser parameters and online sensors, (2) reduce sensor data input with feature extraction (see Fig.\ref{fig:tsfel}), (3) combine available input data to feed our models, (4) train a multi-layer perceptron (MLP) model for roughness prediction, (5) leverage the MLP weights to select most relevant features and, (6) by reducing the dimensionality of features, apply less computationally intensive ML models to predict roughness.}
    \label{fig:pipeline}
    \vspace{-1.4em}
\end{figure*}

The proposed pipeline is depicted in Fig.\ref{fig:pipeline} and consists of four stages defined as follows: (A) Data preprocessing for both laser parameters and sensors signal; (B) Baseline model definition; (C) Feature importance selection; (D) Applicability with ML models.

\vspace{-0.5em}

\vspace{-0.5em}
\subsection{Data preprocessing}
\label{ssec:data_preprocessing}

\begin{figure}[H]
    \vspace{-2.5em}
    \centering
    \includegraphics[width=0.65\textwidth]{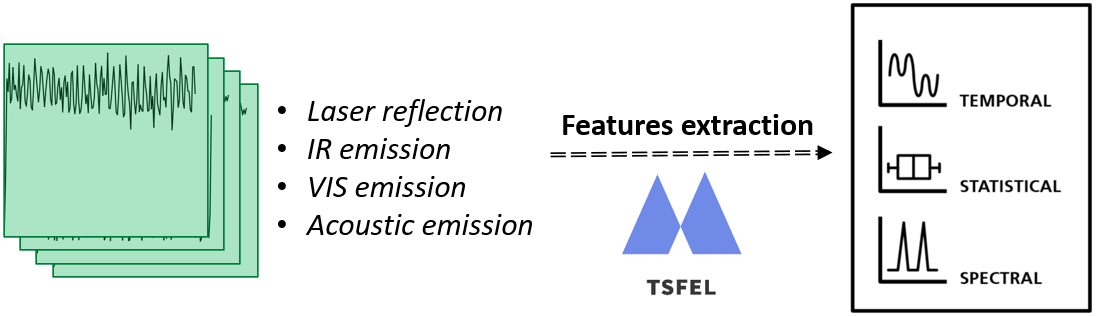}
    \caption{Data preprocessing of the input sensor data that performs feature extraction using TSFEL, extracting characteristics from different domains.}
    \label{fig:tsfel}
    \vspace{-1.5em}
\end{figure}

To transform the information collected by the sensors into useful input data for the model, we used a Python package designed for feature extraction on time series data, the Time Series Feature Extraction Library (TSFEL) \cite{tsfel}. This enabled us to obtain characteristics from the statistical, temporal, and spectral domains of the available signals, as illustrated in Fig.\ref{fig:tsfel}.

As a result of this feature extraction, 389 features were extracted across the three domains for each sensor: 14 from the temporal domain, 40 from the statistical domain, and 335 from the spectral domain.
To optimize real-time monitoring and reduce computational complexity, we specifically chose to extract characteristics solely from the last layer of the laser processing for each sample processed. This layer is expected to contain the most relevant information for predicting final surface roughness.

After extracting these features and assigning them with their corresponding laser parameters, we apply a min-max normalization to help the model better manage the values during training, thereby improving its generalization ability. In this way, each parameter and each feature is normalized based on its minimum and maximum values, confining the values between 0 and 1.

Finally, as a last data preprocessing step to prepare for model training, we apply a training-evaluation split to the dataset, where 80\% is allocated for training and 20\% for evaluation. This allows us to reliably evaluate the model with unseen data from the training process, determining if the model can truly generalize to new data. In addition, a cross-validation with $k=4$ is applied on the training partition, as a resampling procedure to evaluate machine learning models on the limited data samples available.

\vspace{-1em}
\subsection{Training phase}
\label{ssec:training_phase}

\subsubsection{Model architecture.}

The proposed solution leverages the multi-layer perceptron (MLP) \cite{mlppaper}, a model based on neural networks capable of correctly predicting a key property in the subsequent study by establishing non-linear relationships between the different laser parameters and characteristics extracted from the sensor signals.
Its training process consists of a forward propagation algorithm followed by a backpropagation process adjusting the weights of the connections to minimize the loss or error function.\\

\vspace{-2.4em}
\subsubsection{Loss function.}

We used the Mean Squared Error (MSE) to optimize models, a function that measures the error between each element in the input $x$ and target $y$ as follows:
\vspace{-1.7em}

\begin{equation*}
    \mathrm{MSE}=\frac{1}{N} \sum_{i=1}^N\left(y_i-\hat{y}_i\right)^2
\end{equation*}
\vspace{0.2em}
where $N$ is the total number of data points in the dataset, $y_i$ is the ground truth, and $\hat{y}_i$ is the predicted value.

\vspace{-1.4em}
\subsubsection{Hyperparameter optimization.}

In the training phase of the model, we perform a thorough fine-tuning of model parameters to maximize its performance. Traditionally, this fine-tuning process relied on a trial-and-error approach, exploring values known to typically deliver satisfactory results, often employing a brute force grid search. However, in this study, we used the tree-structured Parzen estimator \cite{watanabe2023tree}, a Bayesian optimization algorithm that leverages probabilistic modeling to intelligently guide the search for optimal hyperparameter tuning. This algorithm is the foundation of the Optuna \cite{optuna} library, an open-source hyperparameter optimization framework engineered to automate the hyperparameter search process. Employing this tool, we successfully identify the optimal values for the batch size, learning rate, L1 and L2 regularizations, and determine the most effective architecture for the MLP model. We also explore various optimizers and activation functions. \\

\vspace{-2.6em}
\subsubsection{Feature reduction.}

This project is situated within the context of an industrial production line, where efficiency and precision are of paramount importance. However, achieving real-time implementation requires a significant reduction in the computational costs of the monitoring system. To strike a balance between these two factors, we employ an explainability method to meticulously pinpoint the features that have the most significant impact on the model training process. This approach allows us to streamline the number of parameters involved, thereby reducing computational costs. For increased transparency and interpretability, we aim to achieve a reproducible solution for broader applications, specifically identifying influential features.\\
Therefore, to maximize the feature selection explainability, we leverage the SHapley Additive exPlanations (SHAP) \cite{shap} package, a robust Python library grounded in Game Theory. SHAP assigns a Shapley value to each input feature within the predictive context, quantifying the contribution of the respective feature towards a specific prediction decision. With this method, we aim to add valuable insights into the inner workings of the model.

\vspace*{-3mm}
\section{Experiments}
\label{sec:experiments}

\vspace{-0.5em}
\subsection{Experimental settings}

\vspace{-0.2em}
\subsubsection{Training experiments.}

To rigorously assess the effectiveness of the proposed approach, we conducted a series of experiments aimed at evaluating model performance under various conditions. The experiments were structured to provide comprehensive insights into the adaptability of the model across different laser processing techniques (e.g., milling, grinding, polishing), aiming to optimize its performance as a baseline for future real-time monitoring.
\vspace{-0.8em}
\begin{itemize}
    \item \textbf{Baseline model with laser parameters }as input features. This experiment serves as a baseline to assess model performance without integrating sensor data, focusing on the influence of laser parameters on the predictive capabilities of the model.
    \item \textbf{Model with laser parameters and sensor data} as input, aiming to evaluate the performance improvement achieved by incorporating additional information from sensor measurements.
    \item \textbf{Dimensionality reduction model application.} To optimize the computational requirements, additional ML architectures are compared after reducing the number of input features with the SHAP library based on the best-performing model so far.
\end{itemize}

\vspace{-1.9em}
\subsubsection{Training settings.}

To facilitate a comprehensive comparison of model performance across preprocessing techniques, the training process was carried out in two distinct settings for each approach, individually and in combination. In the individual setting, a separate model was trained for each preprocessing method (i.e., milling, grinding, polishing, die-EDM and wire-EDM), with each preprocessing technique identified by its unique name. In the combined setting, samples from all preprocessing techniques were gathered to train a global model. As for the model architecture, it remained consistent for all training sessions (i.e., different settings) within each approach.

\vspace{-1.5em}
\subsubsection{Evaluation metrics.}

To objectively assess the effectiveness of the proposed models, we used the R-Squared (R2) and Root Mean Squared Error (RMSE) as reference metrics, respectively defined as follows:

\vspace{-1.6em}

\begin{align*}
R^2 &= 1-\frac{\sum\left(y_i-\hat{y}_i\right)^2}{\sum\left(y_i-\bar{y}\right)^2} , & R M S E &= \sqrt{\frac{\sum_{i=1}^N\|y_i-\hat{y}_i\|^2}{N}}
\end{align*}

\vspace{0.2em}

Given $y_i$ as the ground truth, $\hat{y}_i$ as the model predictions, and $N$ as the total number of observations, the model fits better when R2 is closer to 1 and RMSE approaches 0.

\vspace{-1em}
\subsection{Results}
\vspace{-0.5em}

\subsubsection{Baseline model with laser parameters.}

When training the MLP model with the 4 laser parameters and the initial surface roughness, the optimal architecture determined using the Optuna library resulted in an MLP with a single hidden layer composed of three neurons.
As presented in Table \ref{tab:table1}, the results for this experimental setup highlight the necessity for further refinement to ensure robust performance across diverse preprocessing scenarios. While the combined performance is satisfactory when training solely with laser parameters, reaching an R2 metric on the test set of 0.9710 and 0.9463 for milling (a) and combined (f) settings, respectively, its generalizability and adaptability to all preprocessing techniques are not consistent. This observation holds particular significance given that in a real-world scenario, data variability would be substantially higher.

\begin{table}[htbp]
\centering
    \vspace{-0.7em}
    \resizebox{0.7\textwidth}{!}{%
    \begin{tabular}{ccccc}
    & \textbf{Train R2} & \textbf{Test R2} & \textbf{Train RMSE} & \textbf{Test RMSE} \\
    Milling \hspace{5mm} (a)    &     0.9726     &      0.9710    &     1.0786    &    1.0799      \\
    Grinding \hspace{3.5mm} (b)    &    0.4026      &      0.2285    &     0.8585     &    0.8596      \\
    Polishing \hspace{3mm} (c)    &     0.4730     &     0.3749     &     0.1627     &  0.1773        \\
    Die-EDM \hspace{2.5mm} (d)    &     0.9231      &    0.9137      &    0.6145      &   0.6547       \\
    Wire-EDM \hspace{1mm} (e)     &     0.5425     &     0.4771     &     0.3067     &   0.2859       \\
    Combined \hspace{2mm} (f)    &    0.9514      &     0.9463    &     1.0742     &     1.0539     \\ 
    \end{tabular}
    }
    \vspace{0.4em}
\caption{Results across all settings for the model trained with laser parameters only.}
\label{tab:table1}
\vspace{-3em}
\end{table}

By incorporating the information extracted from sensor signals, we aim to improve the model's robustness and observe whether the results show improvement across various preprocessing techniques. In this new setup, the model's input now consists of 1561 input neurons, combining the initial 5 parameters with the 1556 features extracted from sensor data. This change alters the structure of the model compared to the previous configuration, thus resulting in an optimized model architecture of two hidden layers with 443 and 19 neurons, respectively, using the Optuna optimization package. Results obtained with this updated configuration are presented in Table \ref{tab:table2}, comparing results obtained in \cite{zuric2023}, highly sensitive to the preprocessing technique, with ours.


\begin{table}[H]
\centering
    \resizebox{0.75\textwidth}{!}{%
    \begin{tabular}{ccccc}
    & \textbf{Train R2} & \textbf{Test R2} & \textbf{Train RMSE} & \textbf{Test RMSE} \\ 
    \hline
    \multicolumn{5}{c}{\textit{\cite{zuric2023} approach on feature set G}} \\
    Milling \hspace{5mm} (a)    &     0.9881     &     0.9925     &     0.6748     &     0.5560     \\
    Grinding \hspace{3.5mm} (b)  &     0.8757     &     0.6693     &     0.4144     &     0.6746     \\
    Polishing \hspace{3mm} (c)   &     0.9683     &     0.8169     &     0.0377     &     0.1065     \\
    Die-EDM \hspace{2.5mm} (d)   &     0.9928     &     0.9733     &     0.1666     &     0.3749     \\
    Wire-EDM \hspace{1mm} (e)    &     0.7850     &     0.2339     &     0.1863     &     0.5133     \\
    Combined \hspace{2mm} (f)  &     X     &     X     &     X     &     X     \\
    \hline
    \multicolumn{5}{c}{\textit{Our solution}} \\
    Milling \hspace{5mm} (a)    &     0.9993     &      0.9812    &     0.1693    &    0.8682      \\
    Grinding \hspace{3.5mm} (b)    &    0.9866      &      0.8227    &     0.1222     &    0.5180      \\
    Polishing \hspace{3mm} (c)    &     0.9958     &     0.8931     &     0.0146     &  0.0733        \\
    Die-EDM \hspace{2.5mm} (d)    &     0.9997      &    0.9649      &    0.0379      &   0.4177       \\
    Wire-EDM \hspace{1mm} (e)    &     0.9996     &     0.8225     &     0.0094     &   0.1666       \\
    Combined \hspace{2mm} (f)    &    0.9971      &     0.9811    &     0.2628     &     0.6264     \\ 
    \end{tabular}
    }%
    \vspace{0.4em}
\caption{Comparative table between the results obtained in \cite{zuric2023} (above) and ours (below).} 
\label{tab:table2}
\vspace{-3.5em}
\end{table}

\begin{figure*}[hbtp]
\vspace{-1.em}
    \centering
    \subfigure[]{
        \includegraphics[width=0.245\textwidth]{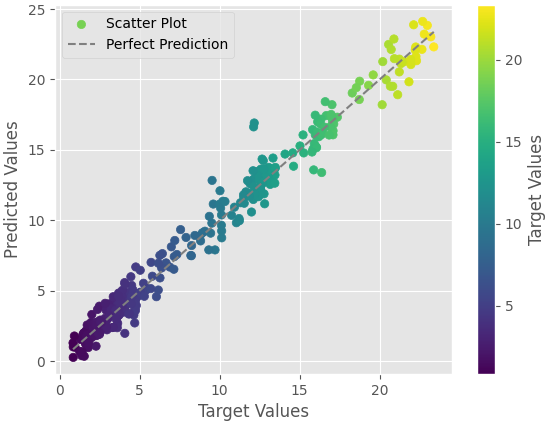} 
        \label{fig:milling_scatter2}
    }
    \subfigure[]{
        \includegraphics[width=0.236\textwidth]{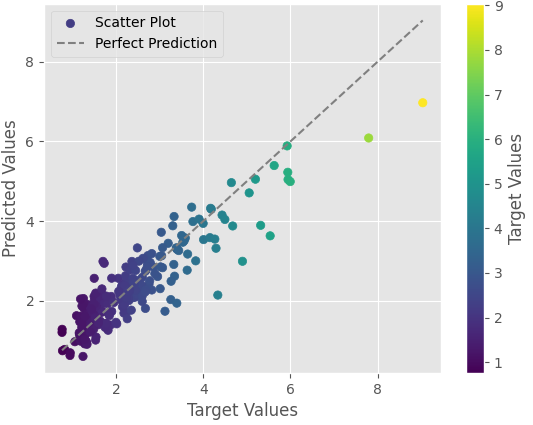} 
        \label{fig:grinding_scatter2}
    }
    \subfigure[]{
        \includegraphics[width=0.245\textwidth]{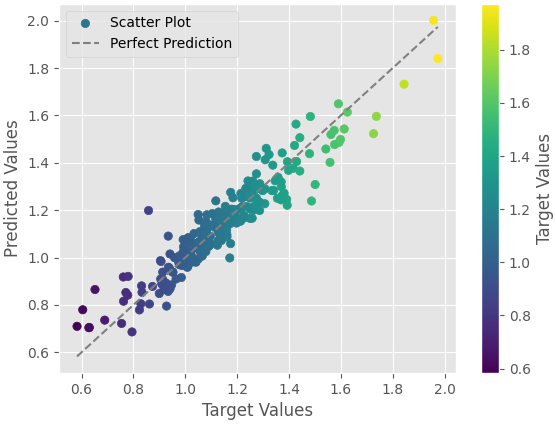} 
        \label{fig:polishing_scatter2}
    }
    \subfigure[]{
        \includegraphics[width=0.236\textwidth]{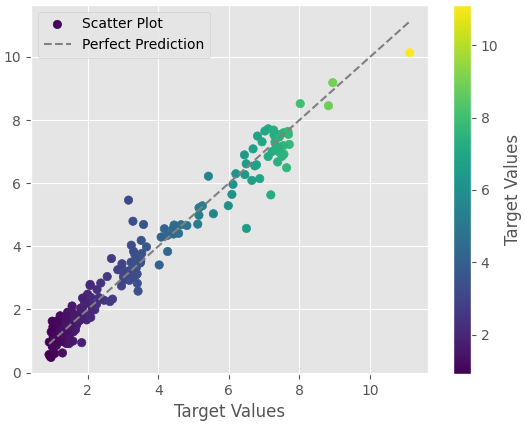} 
        \label{fig:dieedm_scatter2}
    }
    \subfigure[]{
        \includegraphics[width=0.245\textwidth]{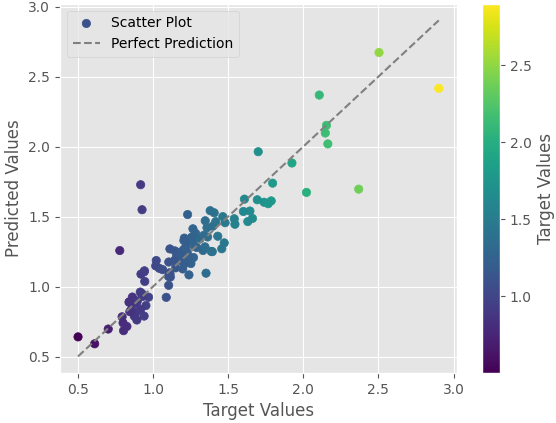} 
        \label{fig:wireedm_scatter2}
    }
    \subfigure[]{
        \includegraphics[width=0.245\textwidth]{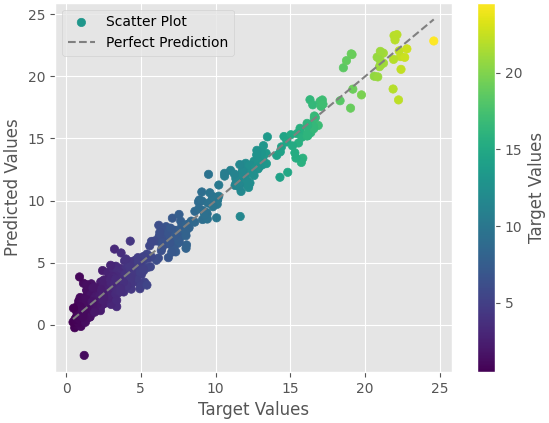} 
        \label{fig:overall_scatter2}
    }
    \vspace{-1.4em}
    \caption{Scatter plots corresponding obtained on the evaluation set, representing data point predictions with respect to ground truth.}
    \label{fig:results2}
    \vspace{-3em}
\end{figure*}

In comparing our results with the model trained with laser parameters only (Table \ref{tab:table1}), incorporating sensor data demonstrates notable improvements in model generalization across all preprocessing techniques. Specifically, previously low results for preprocessing techniques such as grinding (b), polishing (c), and wire-EDM (e) now show a significant improvement, with the R2 metric for grinding (b), for example, increasing from 0.4026 to 0.9866 on the train set and 0.2285 to 0.8227 on the test set. These outcomes underscore the effectiveness of incorporating sensor data in enhancing model performance across diverse preprocessing scenarios. Fig.\ref{fig:results2} shows a scatter plot comparing the ground truth values to the predicted ones for this improved model, where each dot represents an individual data point’s actual and predicted values. The alignment of dots along the diagonal indicates a close match between predicted and actual values, affirming the high performance of the prediction model.

In comparing our approach with the one presented in \cite{zuric2023}, ours shows significant improvement across all preprocessing techniques. Notably, while results for the grinding process (b) in \cite{zuric2023} achieved a test R2 of 0.6693, our approach demonstrates a performance of 0.8227 in the same metric. These comparisons are broken down in Table \ref{tab:table2}.

\vspace{-0.5em}

\vspace{-2em}
\subsubsection{Dimensionality reduction model application.}

The top-performing baseline model was used as a foundation to identify most relevant features with the aim of optimizing the computational requirements of USP monitoring.

\begin{figure}[H]
\vspace{-1em}

    \centering
    \includegraphics[width=\textwidth]{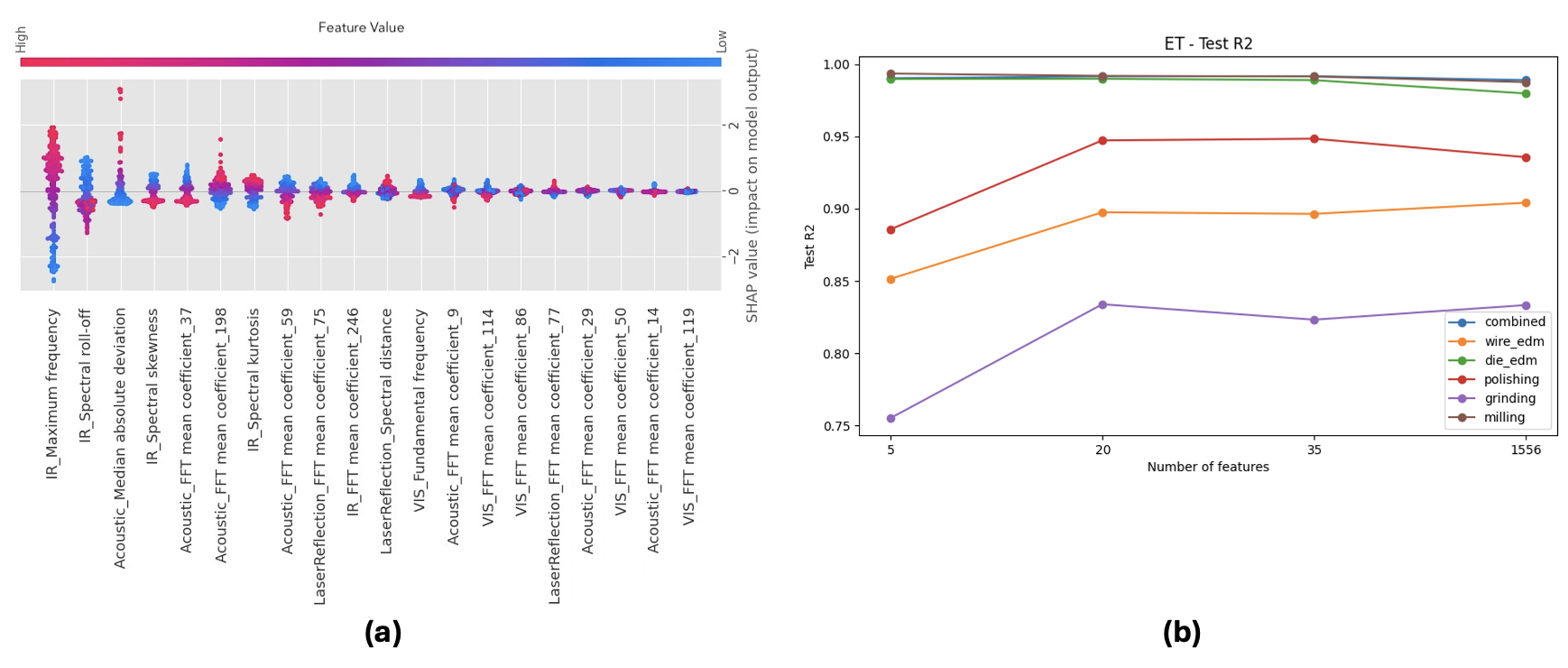}
    \vspace{-1.5em}
    \caption{(a): Example of the 20 selected Shapley feature values obtained for grinding preprocessing. (b): R2 test results on all preprocessing techniques, for different thresholds of selected features extracted with SHAP \cite{shap}. Points on the extreme right of the figure refer to reference results obtained without reducing features.}
    \label{fig:feature_selection}
    \vspace{-1em}
\end{figure}

The Shapley values were extracted with SHAP from the models presented in Table \ref{tab:table2}, to identify which input neurons contributed the most information to the model. Through ablation experiments, we studied the optimal number of sensor features to find a tradeoff between feature reduction and performance. This analysis is depicted in Fig.\ref{fig:feature_selection}-(b), comparing results with the original 1556 sensor features against significantly reduced sets, with 20 features identified as a minimum yielding good metrics. Further reduction in the number of features had a negative impact on model performance for most preprocessing techniques. The obtained Shapley results for these 20 most impacting features are exemplified in Fig.\ref{fig:feature_selection}-(a), using the top-performing base model for the grinding preprocessing.



This substantial reduction in input feature dimensions paves the way for deploying lighter ML models, ultimately aiming to enable future optimized real-time monitoring. Leveraging the 20 selected sensor features, together with the 5 laser parameters and initial surface roughness, we conducted additional trainings using diverse ML models, encompassing extremely randomized trees (ET) \cite{ExtraTrees}, Random Forest (RF) \cite{RF}, Decision Trees (DT) \cite{DecisionTrees}, CatBoost (CB) \cite{CatBoost}, LightGBM (LGBM) \cite{LightGBM} and XGBoost (XGB) \cite{XGB}. The ET reached the best results on all preprocessing techniques except for polishing, where CB slightly outperformed the ET. Results for the top-performing model, i.e., the ET, are detailed in Table \ref{tab:table4}, and demonstrate that lighter ML models trained with considerably reduced sensor features are capable of achieving results similar to those of an MLP trained with all 1556 sensor features.

\vspace{-1.6em}
\begin{table}[H]
\vspace{-1.2em}
\centering
    \resizebox{0.7\textwidth}{!}{%
    \begin{tabular}{ccccc}
    & \textbf{Train R2} & \textbf{Test R2} & \textbf{Train RMSE} & \textbf{Test RMSE} \\
    Milling (a)    &     1.0000     &      0.9956    &     0.0000    &    0.4170      \\
    Grinding (b)    &    1.0000      &      0.8186    &     0.0000     &    0.5237      \\
    Polishing (c)    &     0.9971     &     0.9473     &     0.0119     &  0.0515        \\
    Die-EDM (d)    &     1.0000      &    0.9924      &    0.0000      &   0.1936       \\
    Wire-EDM (e)    &     1.0000     &     0.9243     &     0.0000     &   0.1087       \\
    Combined (f)    &    1.0000      &     0.9934    &     0.0000     &     0.3696     \\ \\
    \end{tabular}
    }
    \vspace{-0.6em}
\caption{Results obtained for each of the processes after selecting the most important features, with the top-performing model (ET).}
\label{tab:table4}
\vspace{-3em}
\end{table}

\vspace{-2em}
\section{Conclusion}
\label{sec:ccl}
\vspace{-1em}
In this paper, we proposed a machine learning approach for laser micromachining optimization that is effective across diverse preprocessing techniques. Our experiments demonstrated that sensors are a valuable source of information in predicting the final surface roughness. To address the impracticality of extracting a large number of features in a real-world setting, we conducted an explainability study to select the most influential features. By strategically reducing the dimensionality and computational cost of the input data, we were able to apply simpler machine learning models such as random forest without compromising performance across preprocessing techniques.
Future work will involve testing this solution in our real-time system and aiming to achieve comparable predictions by focusing on earlier layers of the process.

\vspace{-1em}
\subsubsection{Acknowledgments.} Luis Correas, Miguel Camacho-Sánchez and Laëtitia Launet received funding from Horizon Europe, the European Union’s Framework Programme for Research and Innovation, under Grant Agreement No. 101057457 (METAMORPHA).

\vspace{-1em}
\bibliographystyle{splncs04}
\bibliography{paper}

\end{document}